\def\year{2021}\relax
\documentclass[letterpaper]{article} % DO NOT CHANGE THIS
\usepackage{arxiv}  % DO NOT CHANGE THIS
\usepackage{times}  % DO NOT CHANGE THIS
\usepackage{helvet} % DO NOT CHANGE THIS
\usepackage{courier}  % DO NOT CHANGE THIS
\usepackage[hyphens]{url}  % DO NOT CHANGE THIS
\usepackage{graphicx} % DO NOT CHANGE THIS
\def\UrlFont{\rm}  % DO NOT CHANGE THIS
\usepackage{natbib}  % DO NOT CHANGE THIS AND DO NOT ADD ANY OPTIONS TO IT
\usepackage{caption} % DO NOT CHANGE THIS AND DO NOT ADD ANY OPTIONS TO IT
\usepackage{amsmath,amssymb}
\usepackage{algorithm}
\usepackage{algorithmicx}
\usepackage{algpseudocode}

\usepackage{pifont}
\newcommand{\cmark}{\ding{51}}
\newcommand{\xmark}{\ding{55}}
\usepackage[flushleft]{threeparttable}
\usepackage{array}
\usepackage{multicol}
\usepackage{multirow}
\usepackage{eurosym}

\title{Oral-3D: Reconstructing the 3D Structure of Oral Cavity from Panoramic X-ray}
\author{
    Weinan Song \thanks{Equal Contributions}, 
    Yuan Liang \footnotemark[1], 
    Jiawei Yang, 
    Kun Wang, 
    Lei He
    \\
}
\affiliations{
    University of California, Los Angeles, USA \\
}

\begin{document}
\maketitle

\begin{abstract}
Panoramic X-ray (PX) provides a 2D picture of the patient's mouth in a panoramic view to help dentists observe the invisible disease inside the gum. However, it provides limited 2D information compared with cone-beam computed tomography (CBCT), another dental imaging method that generates a 3D picture of the oral cavity but with more radiation dose and a higher price. Consequently, it is of great interest to reconstruct the 3D structure from a 2D X-ray image, which can greatly explore the application of X-ray imaging in dental surgeries. In this paper, we propose a framework, named \textit{Oral-3D}, to reconstruct the 3D oral cavity from a single PX image and prior information of the dental arch. Specifically, we first train a generative model to learn the cross-dimension transformation from 2D to 3D. Then we restore the shape of the oral cavity with a deformation module with the dental arch curve, which can be obtained simply by taking a photo of the patient's mouth. To be noted, \textit{Oral-3D} can restore both the density of bony tissues and the curved mandible surface. Experimental results show that \textit{Oral-3D} can efficiently and effectively reconstruct the 3D oral structure and show critical information in clinical applications, \textit{e.g.}, tooth pulling and dental implants. To the best of our knowledge, we are the first to explore this domain transformation problem between these two imaging methods.
\end{abstract}

%------------------------------------------------------------------------
\section{Introduction}
% oral diagnosis rely on imaging method

Extra-oral imaging techniques such as PX and CBCT are widely used in dental offices as examination methods before the treatment. Both methods can show detailed bone information, including the tooth, mandible, and maxilla, of the entire oral cavity. However, during the imaging process of PX, the X-ray tube moves around the patient's head and can only take a 2D panoramic picture. This has limited its application in the cases when the disease needs to be located. In comparison, CBCT can reconstruct the whole 3D structure of the lower head with divergent X-rays and provide abundant information about the health condition of oral cavity. Nevertheless, the patient needs to take more radiation dose and pay a higher price during a CBCT scan. We summarize the characteristics of these two imaging methods in Table \ref{tab:Imaging_Compare}. We can see that although CBCT can provide more information in clinical applications \cite{compare_acc}, it generates $39.4\times$ radiation \cite{compare_radiation} and takes $3.7\times$ of the price \cite{compare_fee} on average than PX. This problem is especially evident for those sensitive to the radiation dose and the developing countries where people are unwilling to invest much in dental healthcare. Therefore, it is of great interest to directly reconstruct the 3D structure of the oral cavity from a PX image.

%Panoramic X-ray (PX) and Cone Beam Computed Tomography (CBCT) are two of the most popular imaging methods in digital dentistry. The PX machine takes a 2D picture of all the teeth and the mandible of the patient by moving an X-ray camera around the head. In comparison, the CBCT machine reconstructs the 3D structure of the whole lower head with divergent X-rays. These two methods have been used as general examination methods in many dental clinical applications to observe the diseased tissues that can not be directly observed by eyes. For example, before pulling wisdom teeth, a surgery that most people would have during the lifetime, the dentist generally asks the patient to take a PX examination as the wisdom teeth are usually hidden in the gum. Additionally, in orthodontics, the patient is also required to take the CBCT scan periodically to check the tooth root movement.

\begin{figure}[tp]
    \centering
    \includegraphics[width=0.48\textwidth]{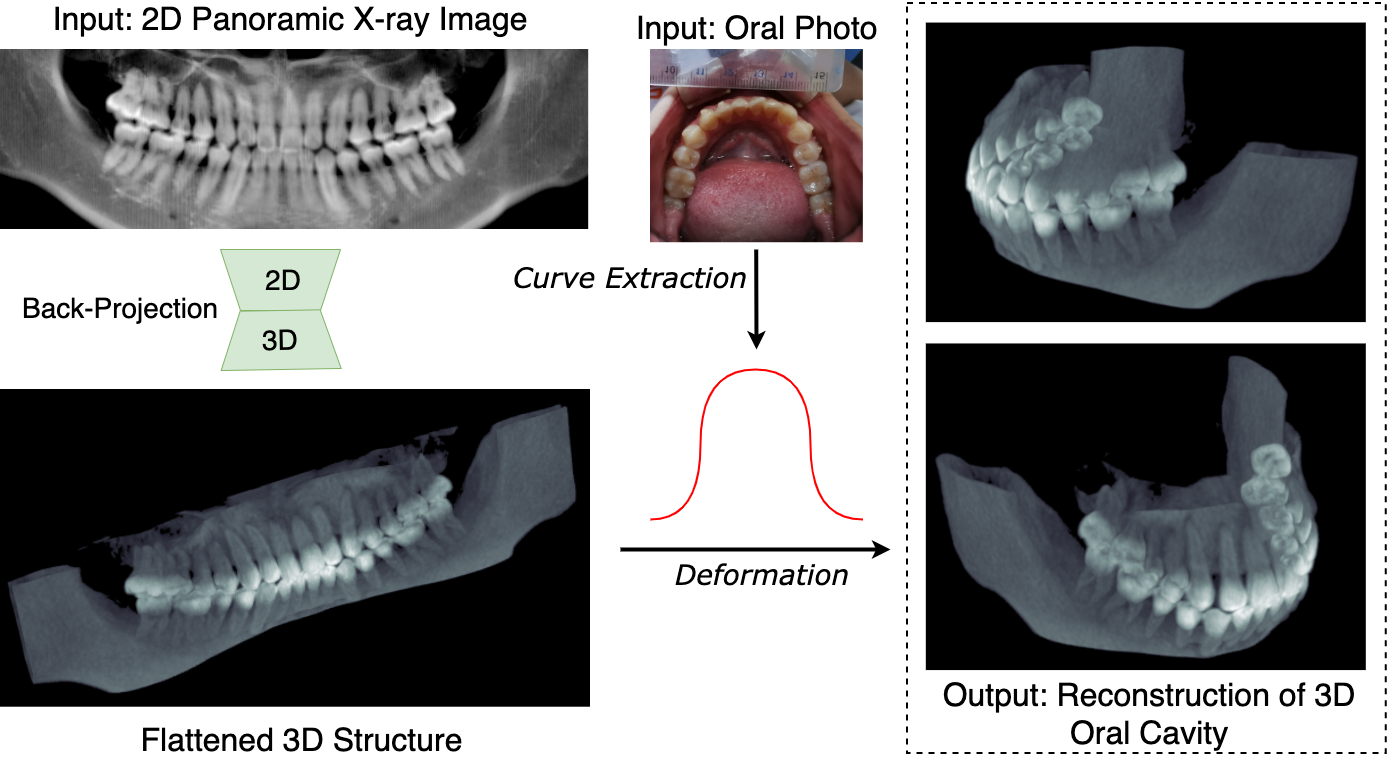}
    \caption{An overview of \textit{Oral-3D}. We first back-project the panoramic image into a flattened 3D image of the oral cavity with a generative network, then we utilize the dental arch information to map the generation result into a curved plane to reconstruct the final 3D structure.}
    \label{fig:Oral-3D}
\end{figure}

\begin{table*}[tp]
  \renewcommand{\arraystretch}{1.3}
  \centering
  \fontsize{8}{10}\selectfont
  \caption{A comparison of CBCT and panoramic X-ray on common dental disease.}
  \label{tab:Imaging_Compare}
  \setlength\tabcolsep{2pt}
  \begin{tabular}{p{1.5cm}<{\centering}p{1.5cm}<{\centering}p{2.6cm}<{\centering}p{2.4cm}<{\centering}p{3cm}<{\centering}p{1.1cm}<{\centering}p{1.1cm}<{\centering}p{1.1cm}<{\centering}p{1.5cm}<{\centering}}
  \hline
  Imaging Method&Dimension&Imaging Cost \cite{compare_fee}&Radiation Dose \cite{compare_radiation}&Diagnostic Accuracy \cite{compare_acc}&Wisdom Tooth&Tooth Decay&Implant Planning&Orthodontics \cr
  \hline
  CBCT &3D &\EUR 184.44 &58.9-1025.4 $\mu$Sv &94.8\% &\cmark &\cmark &\cmark &\cmark \cr
  PX &2D &\EUR49.29 &5.5-22.0 $\mu$Sv &83.3\% &\cmark &\cmark &\xmark &\xmark \cr
  \hline
  \end{tabular}
\end{table*}

 However, it is of great challenge to reconstruct a 3D object from a single 2D image due to the lack of spatial information in the rendering direction. Most works rely on additional information, such as shadow or prior shape of the object, to regularize the reconstruction result. Furthermore, this problem is more difficult for the oral cavity due to the complicated shape of the mandible and detailed density information of the teeth. To overcome such challenges, we propose a two-stage framework, named \textit{Oral-3D}, to generate a high-resolution 3D structure of the oral cavity by decoupling the reconstruction process of shape and density. We first train a generation model to extract detailed density information from the 2D space, then restore the mandible shape with the prior knowledge of the dental arch. Although our method can not totally replace CBCT in the dental examination, we provide a compromise solution to obtain the 3D oral structure when only the PX is available.

% our method
Our work can be summarized as a combination of a single-view reconstruction problem and a cross-modality transformation problem, where the model should recover both the shape and density information of the target object from a single image. We show an overview of \textit{Oral-3D} in Figure \ref{fig:Oral-3D}. At first, we train a generation network to learn the cross-dimension transformation that can back-project the 2D PX image into 3D space, where the depth information of teeth and mandible can be learned automatically from the paired 2D and 3D images. In the second step, we register this generated 3D image, a flattened oral structure, into a curved plane to restore the original shape according to the dental arch. This prior knowledge effectively restricts the shape and location of the 3D structure and can be obtained in many ways, such as by fitting the $\beta$ function with the width and depth of the mouth \cite{beta_curve}. To show the effectiveness of our framework, we first compare \textit{Oral-3D} with other methods on synthesized images generated from a CBCT dataset. Then we evaluate the reconstruction results for some clinical cases to prove the feasibility of our method. Experimental results show that \textit{Oral-3D} can reconstruct the 3D oral structure with high quality from a single panoramic X-ray image and keep the density information simultaneously. In conclusion, we make the following contributions:
\begin{itemize}
    \item We are the first to explore the cross-modality transfer of images in different dimensions for dental imaging by deep learning. In addition to restoring the 3D shape and surface of the bone structure, our model can restore the density information simultaneously, which is of great help for dental diagnosis.
    
    \item We decouple the reconstruction process for density and shape recovery by proposing a deformation module that embeds a flattened 3D image into a curved plane. This has not been addressed in previous research and can significantly improve the reconstruction performance.
    
    \item We propose an automatic method to generate paired 2D and 3D images to train and evaluate the reconstruction models, where \textit{Oral-3D} achieves relatively high performance and can show key features of some typical cases. Meanwhile, we propose a workflow to evaluate our model on a real-world dataset, which indicates the feasibility of clinical applications.
\end{itemize}
\section{Related Work}

\subsection{Deep Learning for Oral Health}
Deep learning has dramatically promoted the computer assistance system for dental healthcare by automatically learning feature representations from large amounts of data. For example, \cite{Toothnet} proposes an automatic method for instance-level segmentation and identification of teeth in the CBCT image. \cite{Cary_detect} trains a deep neural network to detect and diagnose dental caries from periapical radiographic images. \cite{Plaque_class} designs a classification model for red auto-fluorescence plaque images to assist in detecting dental caries and gum diseases. \cite{Dental_transfer} uses transfer learning to classify three different oral diseases for X-ray images. Although these methods have improved oral healthcare service by providing intelligent assistance, the model needs to be trained with annotations on large datasets, which requires both professional knowledge and tedious labour. Compare with these works, our model helps dental healthcare without the supervision of labelled data, where the 3D reconstruction is learned from the latent relationship between 2D and 3D images.

\subsection{Cross-Modality Transfer in Medical Imaging}
\begin{figure}[tp]
    \centering
    \includegraphics[width=0.48\textwidth]{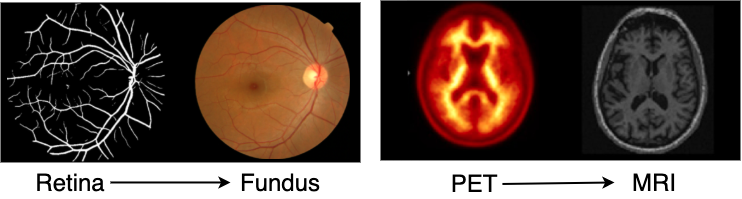}
    \caption{We show some examples of cross-modality transfer for Retina $\rightarrow$ Fundus \cite{Retina2Fundus} and PET $\rightarrow$ MRI \cite{PET2MRI}, where the source image and the target image usually contains consistent physiological structures although in different modalities.}
    \label{fig:related}
\end{figure}

The target of cross-modality transfer is to find a non-linear relationship between medical images in different modalities. It can help reduce the extra acquisition time and additional radiation in a medical examination or provide additional training samples without repetitive annotation work to augment the dataset. Most works take this as a pix-to-pix problem, where the layout and the structure are consistent, but the colour distribution is changed after the transformation between images in different modalities. For example, as shown in Figure \ref{fig:related}, \cite{Retina2Fundus} takes the vessel tree of eyes as a condition to synthesis new images for fundus photography. \cite{PET2MRI} proposes a generation network to produce realistic structural MR images from florbetapir PET images. However, few studies have discussed the cross-modality transfer problem from a lower-dimension image to a higher-dimension one, which is more challenging as the model needs to infer high-dimension information from the lower-dimension image. We only find two works that achieve a similar target to ours. Specifically, \cite{SingleCT} uses an encoder-decoder network to reconstruct 3D skull volumes of 175 mammalian species from 2D cranial X-rays, but the result is subject to too much ambiguity. To improve the visual quality, \cite{X2CT} utilizes biplanar X-rays to extract 3D anatomical structures of body CT with adversarial training and reconstruction constraints. However, our problem is quite different from theirs as the PX image can not be synthesized only from the orthogonal projection over the corresponding CT. Besides, our task is more challenging due to the complicated structure of the oral cavity, where the model is required to restore more details of the teeth and mandible.

\subsection{3D Reconstruction from 2D Image}
Recent work of 3D reconstruction from 2D images can be concluded as two categories: multi-view reconstruction and single-view reconstruction. For the first one, the method generally requires little prior knowledge about the 3D object as the images taken from multiple angles can restrict the reconstruction shape. For example, \cite{RC_Geometric} computes the most probable 3D shape that gives rise to the observed colour information from a series of calibrated 2D images. \cite{3DR2N2} learns the mapping function from arbitrary viewpoints to a 3D occupancy grid with a 3D recurrent neural network. As a comparison, reconstruction from a single-view image usually requires additional information, \textit{e.g.,} prior shape, to inference the object shape. As such, \cite{RC_Position} proposes a unified framework trained with a small amount of pose-annotated images to reconstruct a 3D object. \cite{RC_Shape} takes the adversarially learned shape priors as a regularizer to penalize the reconstruction model. However, the PX image can be either seen as a single-view image taken by a moving camera or a concatenate image blended with multiple views. In this paper, we take our problem as the first kind to decouple the reconstruction process for the bone density and the mandible shape. In the experiment, we also show that this can significantly promote performance over the multi-view reconstruction model both in quality and quantity.
\section{Method}

\begin{figure*}[tp]
    \centering
    \includegraphics[width=0.95\textwidth]{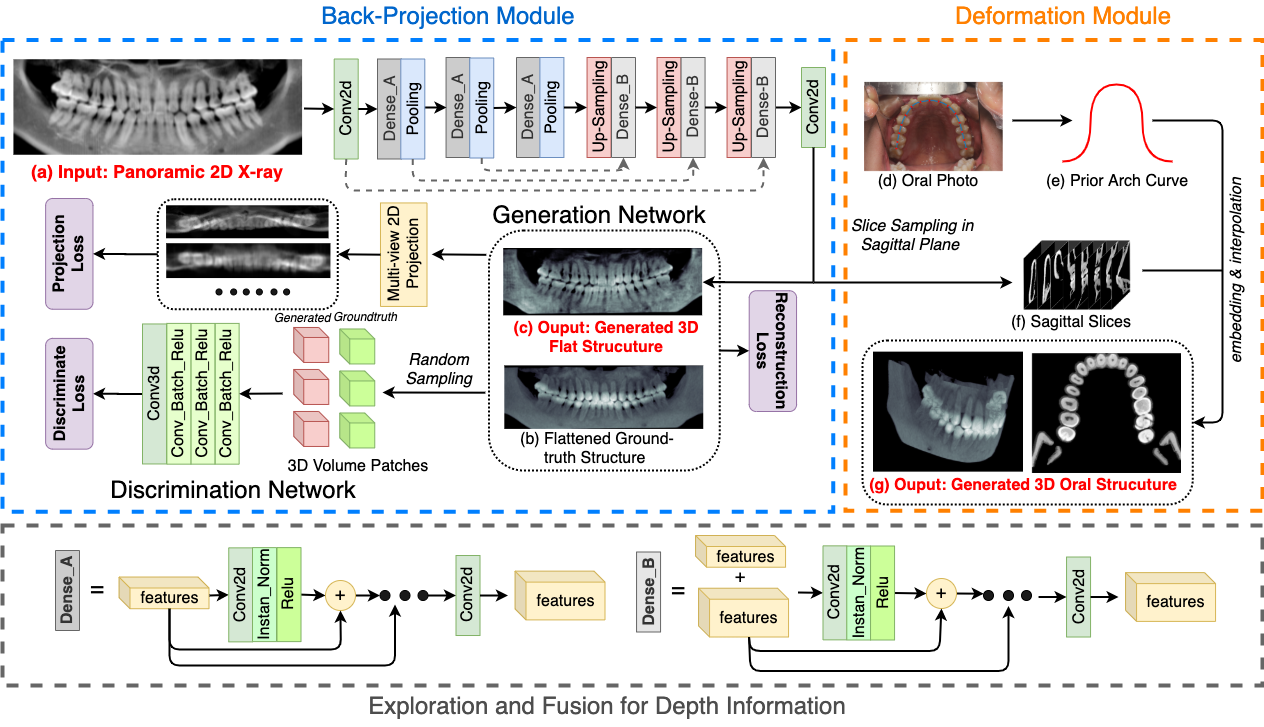}
    \caption{Our framework consists of two modules to decouple the recovery of bone density and the mandible shape. The back-projection module utilizes a generation network to restore the 3D density information from the 2D space, and the deformation module transforms the flattened 3D image into a curve plane according to the prior knowledge in the dental arch.}
    \label{fig:Model}
\end{figure*}

In this section, we introduce our framework that reconstructs a high-resolution 3D oral cavity from a 2D PX image. We choose to break this problem into two stages to recover more details of the bone density. We show the structure of \textit{Oral-3D}, which consists of a back-projection module and a deformation module in Figure \ref{fig:Model}. The back-projection module develops from generative adversarial networks (GAN \cite{GAN}), where the generator is trained to learn a back-projection transformation by exploring the depth information contained in the X-ray image. The deformation module takes in the generated 3D image (Image $c$) from the back-projection module and the dental arch (Image $e$) to restore the curved shape of the mandible.

\subsection{Back-Projection Module}
GANs have proved to be an effective model to learn latent data distribution by training the generator $G$ and the discriminator $D$ in an adversarial way. The generator learns to output a fake image from a random vector to deceive the discriminator, while the discriminator tries to distinguish sampling data between real and fake images. As we aim to generate the consistent 3D content from the semantic information of the panoramic X-ray image, we utilize conditional GANs \cite{cGAN} as the generative model to learn the back-projection transformation.

\paragraph{\textbf{Objective Function}}
To improve the generation quality and guarantee the stable training process, we use LSGAN \cite{LSGAN} as the keystone to train the generator and discriminator. The adversarial loss can be summarized as:
\begin{equation}
\label{loss}
\begin{aligned}
Loss_{D} =& \mathbb{E}_{y}\left[(D(y)-1)^{2}\right] + \mathbb{E}_{x} \left[D(G(x))^{2} \right] \\
Loss_{G} =& \mathbb{E}_{x} \left[ D(G(x))-1)^{2} \right],
\end{aligned}
\end{equation}
where \textit{x} is the PX image and \textit{y} is the flattened oral structure.

To maintain the structural consistency of the input and generation result, we also introduce the reconstruction loss and projection loss to improve the generation quality. These proposed loss functions can bring voxel-wise and plane-wise regularization to the generation network, which can be defined as:
\begin{equation}
\label{loss_G}
\begin{aligned}
Loss_{R} =& \mathbb{E}_{x,y} \left[ (y-G(x))^2 \right] \\
Loss_{P} =& \mathbb{E}_{x,y} \left[ (P(y)-P(G(x)))^2 \right],
\end{aligned}
\end{equation}
where the function $P()$ is achieved by orthogonal projections along each dimension of the generated 3D image. In summary, the total optimization problem can be concluded as:
\begin{equation}
\label{loss_GAN}
\begin{aligned}
D^{*} =& \arg \min_{D} Loss_{D} \\
G^{*} =& \arg \min_{G} \lambda_{1} \cdot Loss_{G} + \lambda_{2} \cdot Loss_{R}+ \lambda_{3} \cdot Loss_{P}.
\end{aligned}
\end{equation}

\paragraph{\textbf{Generator}}
During the X-ray imaging, the depth information can be reflected in the absorption of radiation through the bone. Therefore it is reasonable to extract the thickness of the tooth and the mandible from a PX image. Then the objective for the generator is to find a cross-dimension transformation $G$ from 2D to 3D, which can be denoted as:
\begin{equation}
    \label{eq:template_generation}
    {G}:{I^{2D}_{H \times W}} \to {I^{3D}_{H \times  W \times D}},
\end{equation}
where $I^{2D}$ is the PX image with a size of ${H \times W}$ and $I^{3D}$ is the flattened 3D structure with a size of ${H \times  W \times D}$. In this paper, we utilize 2D convolution to retrieve the latent depth information. The 3D information is embedded into different channels of feature maps. As shown in Fig. \ref{fig:Model}, the encoding network decreases the resolution of feature maps but increases the number of feature channels, while the decoding network increases the resolution to generate a 3D object. The output voxel value is restricted to $(-1, 1)$ with a $tanh$ layer at the end.

\paragraph{\textbf{Dense Block}}
Dense connections \cite{DenseNet} have shown compelling advantages for feature extraction in deep neural networks. This architecture is especially efficient in forwarding 3D information as each channel of the output has a direct connection with intermediate feature maps. In the projection module, we utilize two kinds of dense blocks, noted as A and B, to extract depth information from the X-ray image. As shown at the bottom of Figure \ref{fig:Model}, the dense block A explores the depth information by increasing the channel number of feature maps. In contrast, the dense block B fuses feature maps from the $up-sampling layer$ and the skip-connections but maintain the number of channels to forward the depth information. In the end, the number of stacked features in the output is equal to the depth of the generated 3D image.

\paragraph{\textbf{Discriminator}}
The discriminator has been frequently used in many generative models to improve the generation quality by introducing an instance-level loss. In the back-projection module, we adopt a patch discriminator introduced by \cite{Img_Translate} to improve the generation quality of tooth edges by learning high-frequency structures in the flattened 3D structure.  We set the patch size as ${70\times70\times70}$ and follow a similar structure in \cite{Img_Translate} but replace 2D convolution with 3D. The discrimination network ends with a $Sigmoid$ layer to predict the probability of the samples belonging to the real image. To be noted, we sample the same number of 3D patches at the same position from the paired of 3D images when training the discriminator.

\subsection{Deformation Module}
With the generation of a 3D image from the back-projection module, the deformation model maps the flattened 3D structure into the curved space according to the arch curve to output the final reconstruction object. As shown in the right part of Figure \ref{fig:Model}, we propose a registration algorithm that can best restore the shape of the oral cavity and keep the recovered density information. We first sample the generated 3D image (Image $c$) into slices (Image $f$) in the sagittal plane, then interpolate these slices along the dental arch curve (Image $e$). To achieve this, we sample a number of points from the curve with equal distance and embed the slices into the curve. In the end, we interpolate the voxels between the neighbouring slices to output a smooth 3D image (Image $g$). For computation convenience, we combine these steps together and conclude it in Algorithm \ref{alg:registration}, where we assume that the generated 3D image and the bone model has the same height of $H$.

\begin{algorithm}[tp]
\caption{Embedding and Interpolation}
\label{alg:registration}
\begin{algorithmic}[1]

\Function {register}{$Slices, W_{3D}, D_{3D}, curve$}
\State $W, H, D \gets $ \Call{shape}{$Slices$}
\State $OralImage \gets $ \Call{zeros}{$W_{3D}, H, D_{3D}$} 
\State $SamplePoints \gets $ \Call{sample}{$curve$}
\For{$i=0;i<W_{3D};i++$}
    \For{$j=0;j<D_{3D};j++$}
        \State $id, dist \gets$ \Call{dist}{$(i, j), SamplePoints$}
        \State $Slice \gets Slices[id, :, :]$
        \State $Slice \gets $ \Call{interpolate}{$Slice, dist$}
        \State $OralImage[i, :, j] \gets Slice$ 
    \EndFor
\EndFor
\State \Return $OralImage$
\EndFunction
\end{algorithmic}
\end{algorithm}

\section{Experiment}

\subsection{Dataset}
As grouped data of PX image, dental arch shape, and 3D oral structure of the same patient, especially in the same period, is hard to find, we first use synthesized data to evaluate the performance. We collect 100 CBCT scans from a major stomatological hospital in China and re-sample these 3D images into a size of $288\times256\times160$. The dataset is finally normalized into a range of $(-1, 1)$ and split into a ratio of $3:1:1$ for training, validation, and testing.

\begin{figure}[tp]
    \centering
    \includegraphics[width=0.45\textwidth]{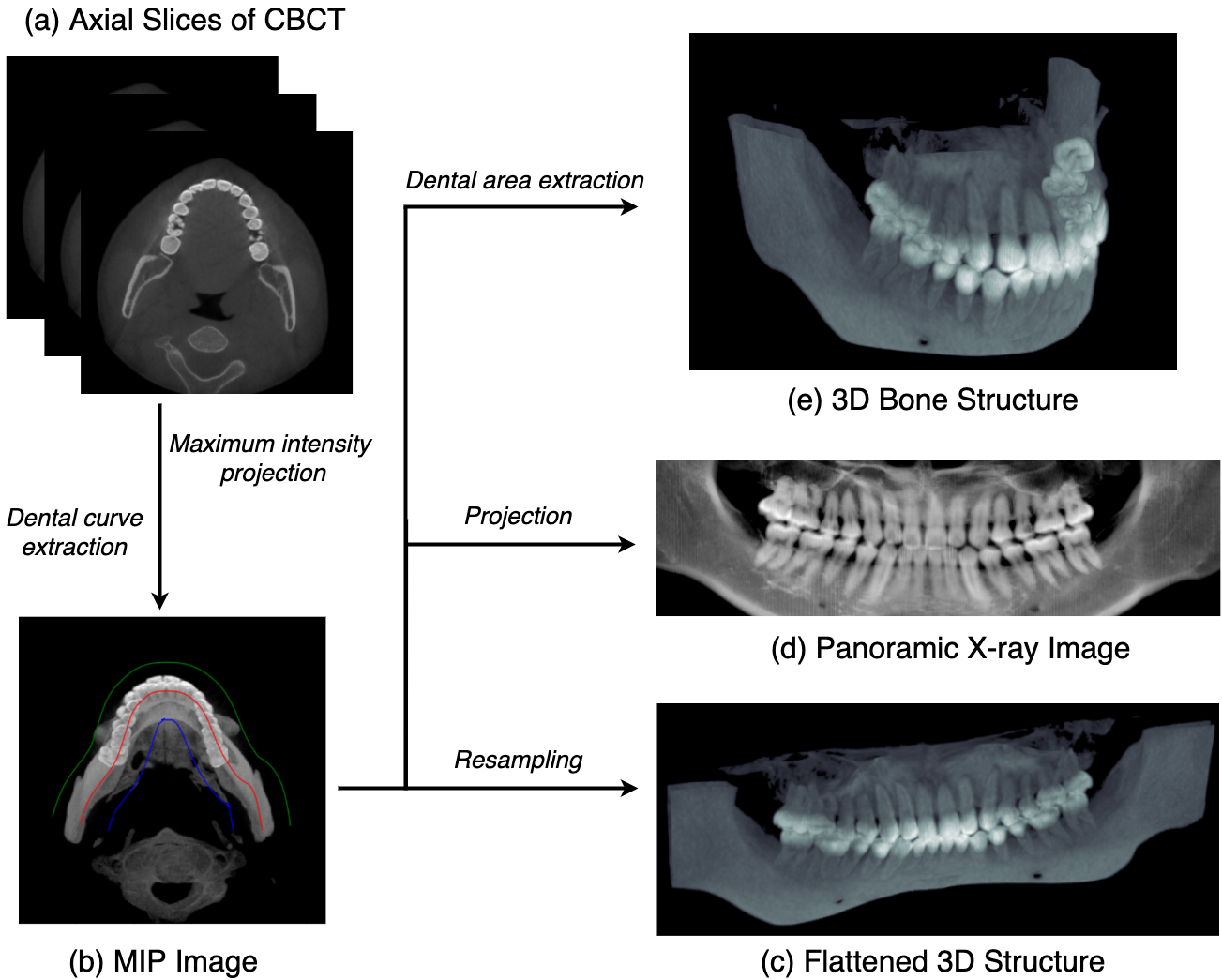}
    \caption{An overview of generating paired data for 3D oral structure and 2D panoramic X-ray is shown in this picture. We first get the MIP image from the CBCT scan to obtain the dental arch curve (red), and boundaries of the dental area (blue and green). Then we obtain the flattened oral structure, PX image, and the 3D oral structure by re-sampling, projection, and extraction, respectively.
    }
    \label{fig:preprocess}
\end{figure}

An overview of preparing the synthesized data can be seen in Figure \ref{fig:preprocess}. We first obtain a 2D image in the axial plane by maximum intensity projection (MIP) (Image $b$) over the CBCT slices (Image $a$). Then we obtain the dental curve with a similar method as in \cite{Panoramic_Generation} to estimate the curve function and boundaries of the dental arch. To generate the PX image (Image $d$), we simulate projection with the Beer-Lambert absorption-only model along the arch curve. This imaging process is similar to the way for a real PX machine, where the manufacturer usually improves the imaging quality by designing a trajectory of the camera to fit the mandible shape. Finally, we extract the 3D oral structure (Image $e$) by removing the unrelated tissues with the boundaries and generate the flattened 3D structure (Image $c$) by re-sampling along the arch curve.

\subsection{Evaluation Metrics}
\begin{itemize}
    \item \textbf{PSNR:} Peak signal-to-noise ratio (PSNR) is often used to measure the difference between two signals. Compared with mean squared error, PSNR can be normalized by the signal range and expressed in terms of the logarithmic decibel scale. We take this to measure the density recovery of our models.
    
    \item \textbf{Dice:} In order to reflect the deformation of the reconstruction, we use dice coefficient between our reconstruction results and the groundtruth in a volume level of the oral cavity. The 3D volume of the oral cavity is obtained by setting a threshold (\textit{e.g.,}$-0.8$ over the reconstruction result.

    \item \textbf{SSIM:} We use the structure similarity index (SSIM) \cite{SSIM} as the key criterion to quantify the performance of density recovery.SSIM considers the brightness, contrast and structure information at the same time and can match better the subjective evaluation of humans. It can effectively indicate the reconstruction quality and is widely used in other similar works, such as \cite{X2CT}.
    
    \item \textbf{Overall:} To combine these three metrics together, we also define a score $S = (PSNR/20 + Dice + SSIM) /3 $ to compare the overall performance of the 3D reconstruction.
    
\end{itemize}

\begin{figure*}[tp]
    \centering
    \includegraphics[width=\textwidth]{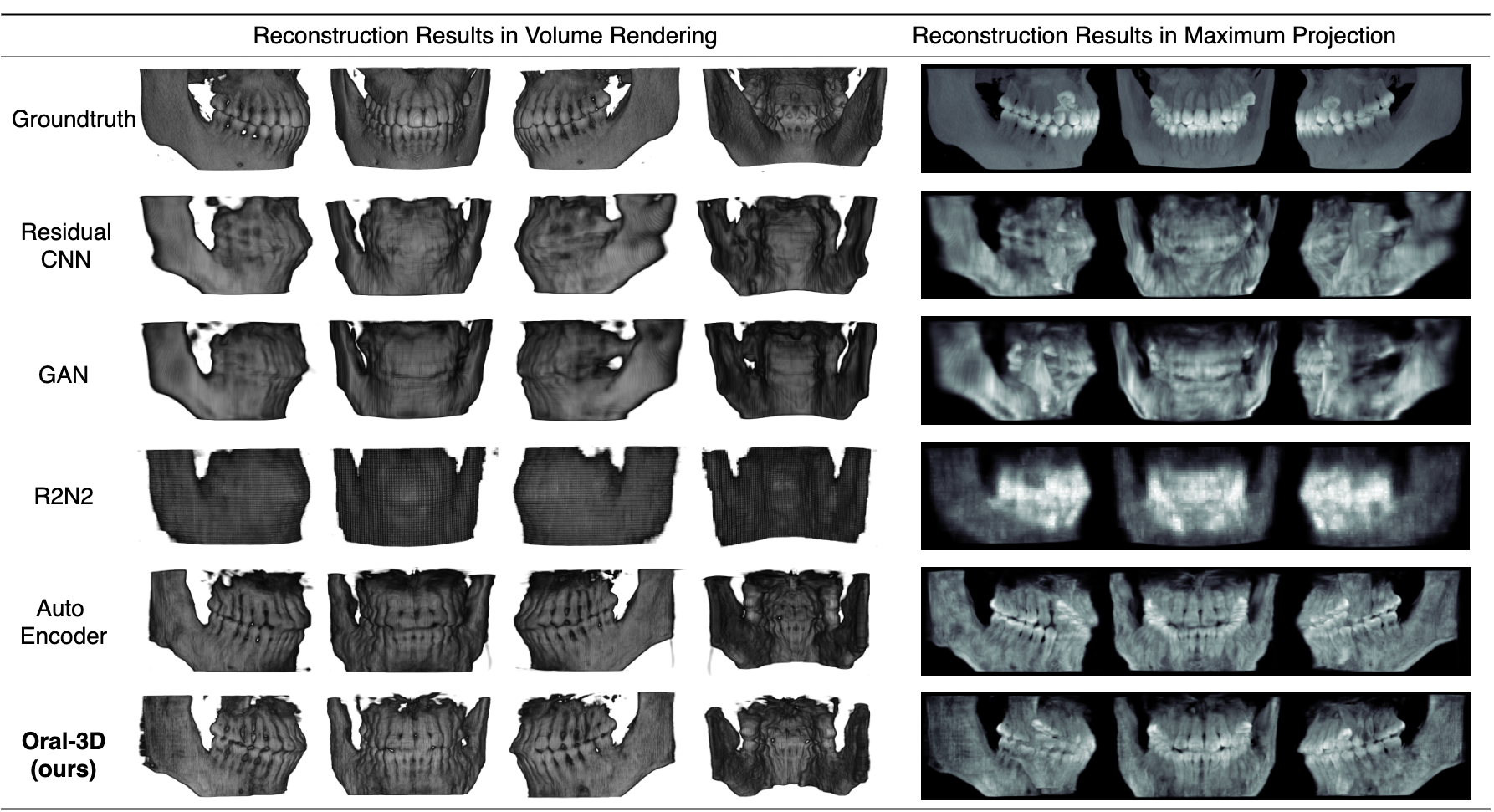}
    \caption{We show the qualitative comparison from different views and rendering ways in this picture. We can see that our method generates the best results with more detailed density and a more sharp surface.}
    \label{fig:compare_shape}
\end{figure*}

\begin{figure}[tp]
    \centering
    \includegraphics[width=0.48\textwidth]{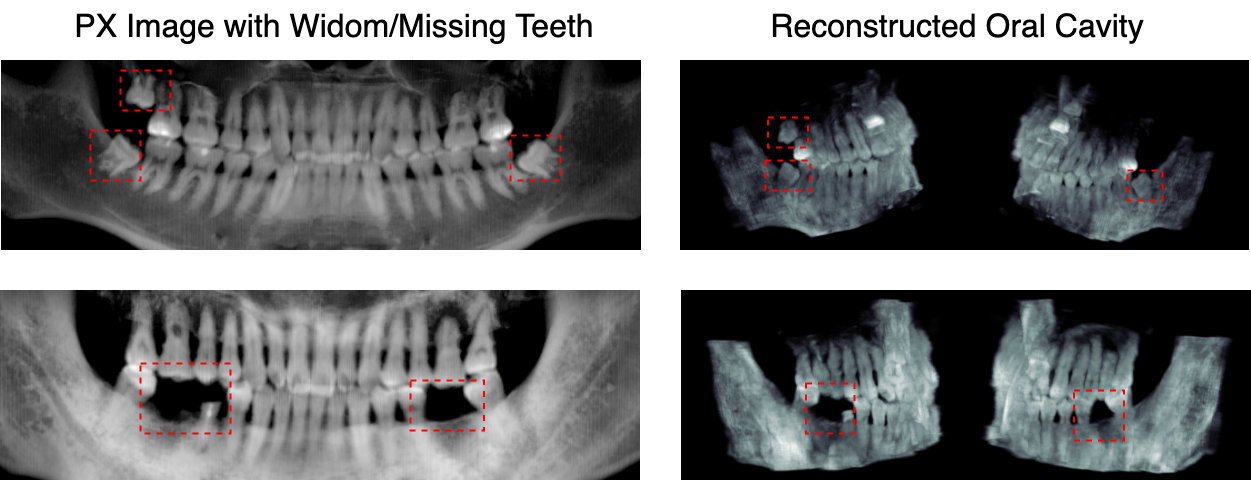}
    \caption{We show reconstruction results for patients with wisdom/missing teeth and mark the key features with red bounding boxes. We can see that our method can accurately locate these positions, which can be an important reference during the surgery.}
    \label{fig:result_disease}
\end{figure}

\subsection{Comparison Models}
To show the effectiveness and efficiency of Oral-3D, we also compare our framework with other models that work on a similar problem:
\begin{itemize}
    \item \textbf{Residual CNN:} An encoder-decoder network that has been introduced in \cite{SingleCT} to reconstruct the 3D model with a single X-ray.
    \item \textbf{GAN:} A generative model based on \cite{GAN} that takes the Res-CNN as the backbone for generator with reconstruction loss and the same discriminator as \textit{Oral-3D}.
    \item \textbf{R2N2:} We transform our task into a multi-view reconstruction problem to train R2N2 \cite{3DR2N2} by taking the PX image as a composition of X-ray image taken from three different views.
    \item \textbf{Auto Encoder:} We remove the discriminative network in \textit{Oral-3D} and keep the encoder-decoder network only in the back-projection module.
\end{itemize}

\subsection{Training}
All the experiment are trained by Adam optimizer \cite{adam} with a batch size of 1 for 300 epochs. The learning rate starts at $1\times10^{-3}$ and decreases 10 times every 50 epochs. We use the validation data as the stop criterion, and all models converge after 300 epochs. For adversarial networks, i.e. \textit{Oral-3D} and GAN, we introduce the discriminative network after 100 epochs to alleviate the influence of discrimination loss at the beginning.
\section{Results}

\begin{figure*}[h]
    \centering
    \includegraphics[width=\textwidth]{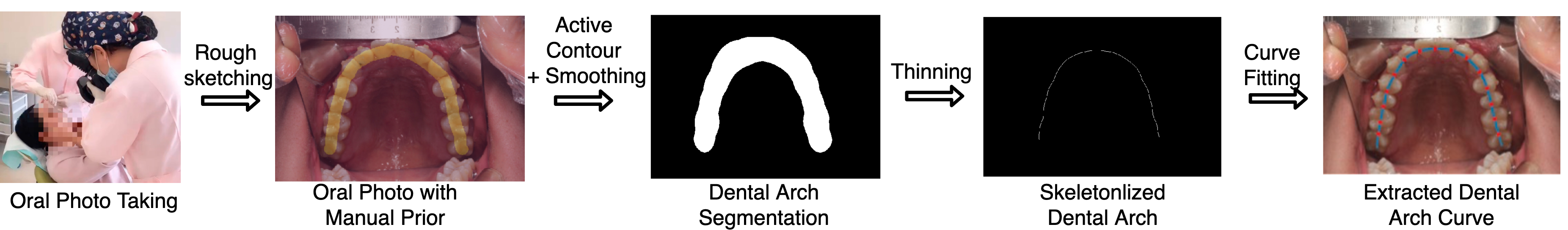}
    \caption{We show a workflow to apply \textit{Oral-3D} to obtain the dental arch curve in real-world applications in this picture. We first take a picture of the patient's mouth and segment then dental area semi-automatically. Then we use a cubic function to the fit points sampled from the skeletonized image of the binary mask.}
    \label{fig:dental_app}
\end{figure*}

\begin{table*}[h]
    \small
    \renewcommand{\arraystretch}{1.2}
    \caption{Quantitative Evaluation of 3D Reconstruction}
    \label{tab:comapre_tab}
    \centering
    \setlength\tabcolsep{2pt}
    \begin{tabular}{p{4cm}<{\centering}p{1cm}<{\centering}p{1cm}<{\centering}p{1cm}<{\centering}p{2.5cm}<{\centering}p{2.5cm}<{\centering}p{2.5cm}<{\centering}p{2cm}<{\centering}}
    \hline
    Method&View&Prior&D-Net&PSNR (dB)&SSIM (\%)&Dice (\%)&Overall\cr
    \hline
    Residual CNN&1&No&No&17.46$\pm$9.58&72.90$\pm$2.09&57.95$\pm$7.43&73.54\cr
    GAN&1&No&Yes&17.71$\pm$1.04&69.96$\pm$1.91&57.80$\pm$7.76&73.78\cr
    R2N2&3&No&No&18.06$\pm$0.94&71.94$\pm$1.36&57.71$\pm$6.52&73.32\cr
    Oral-3D (Auto-Encoder)&1&Yes&No&19.04$\pm$0.85&76.78$\pm$1.65&69.68$\pm$4.98&80.56\cr
    Oral-3D (GAN) &1&Yes&Yes&\textbf{19.22$\pm$0.83}&\textbf{78.27$\pm$1.74}&\textbf{71.28$\pm$4.69}&\textbf{81.89}\cr
    \hline
    \end{tabular}
\end{table*}

\begin{table}[tp]
    \small
    \renewcommand{\arraystretch}{1.2}
    \caption{Evaluation results of different combination of discrimination loss (DL), reconstruction loss (RL), and projection loss (PL).}
    \label{tab:comapre_loss}
    \centering
    \setlength\tabcolsep{2pt}
    \begin{tabular}{p{1cm}<{\centering}p{1.2cm}<{\centering}p{1.8cm}<{\centering}p{1.8cm}<{\centering}p{2cm}<{\centering}}
    \hline
    &DL only&DL+PL&DL+RL&DL+RL+PL\cr
    \hline
    PSNR & 8.06 &18.06(+10.00) &19.14(+11.08) &\textbf{19.22(+11.16)} \cr
    SSIM & 46.61 &73.02(+26.41) &\textbf{78.41(+31.80)} &78.27(+31.66) \cr
    Dice & 35.50 &64.53(+29.03) &70.89(+35.39) &\textbf{71.28(+35.78)} \cr
    Overall & 40.79 &75.95(+35.16) &81.66(+40.87) &\textbf{81.89(+41.10)} \cr
    \hline
    \end{tabular}
\end{table}

In this section, we evaluate the reconstruction performance of \textit{Oral-3D} from different different perspectives. We first compare \textit{Oral-3D} with other methods qualitatively and quantitatively. Then we show the results of special cases for some common dental applications. In the end, we do clinical trails by evaluating our method on real-word images.

% Exp1
\subsection{Comparison with Other Methods}
We first show the 3D bone structure in two rendering ways as in Figure \ref{fig:compare_shape}, where the volume rendering can show the reconstructed surface and the maximum projection can indicate the restored density information. Then we summarize the evaluation metrics in Table \ref{tab:comapre_tab} to compare with other methods. We can see that \textit{Oral-3D} has the best performance over other models. Comparing \textit{Oral-3D} and Auto Encoder with the Residual CNN and GAN, we can see the importance of decoupling the back-projection and deformation process. To be noted, R2N2 achieves the worst performance, where the model only learns the shape of the oral cavity but loses details of teeth. This has indicated the defect when converting the PX image as a collection of multi-view images. Additionally, we see that Auto Encoder has the closest performance to \textit{Oral}, although the latter has a more clear surface. This has proved the promotion brought by the adversarial loss.

% Exp3
\subsection{Identification of Wisdom/Missing Teeth}
In this paragraph, we show two of the most common cases in dental healthcare, \textit{e.g.}, dental implants and tooth pulling, to see if \textit{Oral-3D} can provide dentist useful reference. Both cases require to locate the operation location before the surgery. In the first row of Figure \ref{fig:result_disease}, three wisdom teeth can be seen clearly on both sides in PX. These features also present in the two sides of the reconstruction results. In the second row, the patient misses two teeth on both sides of the mandible. While the missing place can also be located accurately in the reconstruction image.

% Exp4
\subsection{Ablation Study}
To reveal the factors that influence the reconstruction quality of the generation network, we also do an ablation by changing the combination of the loss functions. As shown in Table \ref{tab:comapre_loss}, we see that the model shows the worst performance if trained only with the adversarial loss. This is mainly because the adversarial loss can not bring voxel-wise optimization. We can also see that the major improvement comes from the reconstruction loss, while the projection loss brings much less promotion, especially when trained with the reconstruction loss together. This is also reasonable as the reconstruction loss can supervise the generation network to learn more detailed information.

\begin{table}[tp]
    \small
    \renewcommand{\arraystretch}{1.2}
    \caption{Evaluation results on real-world images}
    \label{tab:comapre_real}
    \centering
    \setlength\tabcolsep{2pt}
    \begin{tabular}{p{1.6cm}<{\centering}p{1.8cm}<{\centering}p{1.8cm}<{\centering}p{1.8cm}<{\centering}}
    \hline
    Dataset&PSNR&SSIM&Dice\cr
    \hline
    Real &17.36$\pm$0.70 &69.30$\pm$2.03&71.44$\pm$3.66\cr
    Synthesized &19.22$\pm$0.83 &78.27$\pm$1.74&71.28$\pm$4.69\cr
    \hline
    \end{tabular}
\end{table}

\begin{figure}[tp]
    \centering
    \includegraphics[width=0.45\textwidth]{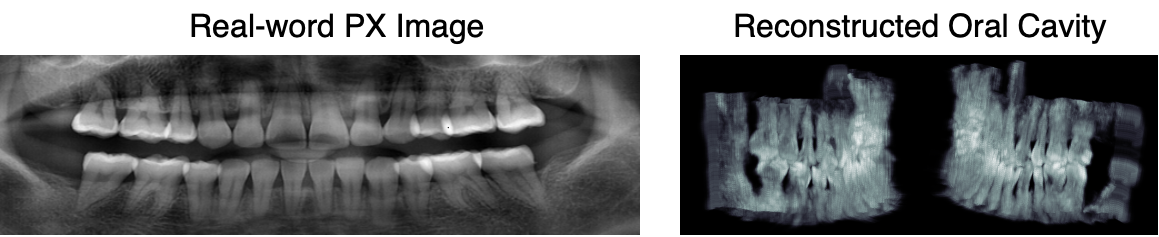}
    \caption{Although the quality decreases in density details for real-word PX, we can still identify each tooth in the reconstruction result.}
    \label{fig:result_real}
\end{figure}

% Exp5
\subsection{Clinical Trials}
In the end, we evaluate \textit{Oral-3D} on real-world data from 6 patients. The workflow of collecting dental arch information is shown in Figure \ref{fig:dental_app}. We use cycleGAN \cite{CycleGAN} to alleviate the colour variance between the training and testing PX images. As shown in Table \ref{tab:comapre_real}, the drop mainly comes from the PSNR and SSIM, which is because the colour variance in different CBCT machines. From Figure \ref{fig:result_real} we can that although the quality decreases in density details, we can still identify each tooth in the reconstruction result.
%------------------------------------------------------------------------
\section{Conclusion}
In this paper, we propose a two-stage framework to reconstruct the 3D structure of the oral cavity from a single 2D PX image, where individual shape information of the dental arch is provided as prior knowledge. We first utilize a generative model to back-project the 2D image into 3D space, then deform the generated 3D image into a curved plane to restore the oral shape. We first use synthesized data to compare with different methods, then evaluate the model with real-world data to see the feasibility in clinical applications. Experimental results show that our model can recover both the shape and the density information in high resolution. We hope this work can help improve dental healthcare from a novel attitude.
%------------------------------------------------------------------------
\section{Acknowledgement}
We thank Dr. Liang Chengwen, Dr. Wangbin, and Dr. Wang Weiqian for collecting data.
\clearpage

\bibliography{reference.bib}

\end{document}